\documentclass[fleqn,usenatbib]{mnras}

\usepackage{newtxtext,newtxmath}

\usepackage[T1]{fontenc}

\DeclareRobustCommand{\VAN}[3]{#2}
\let\VANthebibliography\thebibliography
\def\thebibliography{\DeclareRobustCommand{\VAN}[3]{##3}\VANthebibliography}

\usepackage{graphicx}	
\usepackage{amsmath}	

\title[Purple is the new green: biopigments and spectra of Earth-like purple worlds]{Purple is the new green: biopigments and spectra of Earth-like purple
worlds}

\author[L. F. Coelho et al.]{
Lígia Fonseca Coelho,$^{1,2}$\thanks{E-mail: lc992@cornell.edu}
Lisa Kaltenegger,$^{1,2}$
Stephen Zinder,$^{2,3}$
William Philpot,$^{2,4}$
Taylor L. Price,$^{5}$
\newauthor and Trinity L. Hamilton $^{5,6}$
\\
$^{1}$Department of Astronomy, Cornell University, Space Sciences Building 404, Ithaca, NY 14850, USA\\
$^{2}$Carl Sagan Institute, Cornell University, Space Science Building 311, Ithaca, NY 14850, USA\\
$^{3}$Department of Microbiology, Cornell University, Ithaca, NY 14853, USA\\
$^{4}$School of Civil and Environmental Engineering, Cornell University, Ithaca, NY 14853, USA\\
$^{5}$Department of Plant and Microbial Biology, University of Minnesota, St. Paul, MN 55108, USA\\
$^{6}$The Biotechnology Institute, University of Minnesota, St. Paul, MN 55108, USA\\
}

\date{Accepted 2024 February 16. Received 2024 February 15; in original form 2023 September 30}

\pubyear{2024}

\begin{document}
\label{firstpage}
\pagerange{\pageref{firstpage}--\pageref{lastpage}}
\maketitle

\begin{abstract}
With more than 5500 detected exoplanets, the search for life is entering a new era. Using life on Earth as our guide, we look beyond green landscapes to expand our ability to detect signs of surface life on other worlds. While oxygenic photosynthesis gives rise to modern green landscapes, bacteriochlorophyll-based anoxygenic phototrophs can also colour their habitats and could dominate a much wider range of environments on Earth-like exoplanets. Here, we characterize the reflectance spectra of a collection of purple sulfur and purple non-sulfur bacteria from a variety of anoxic and oxic environments. We present models for Earth-like planets where purple bacteria dominate the surface and show the impact of their signatures on the reflectance spectra of terrestrial exoplanets. Our research provides a new resource to guide the detection of purple bacteria and impro ves our chances of detecting life on exoplanets with upcoming telescopes. Our biological pigment data base for purple bacteria and the high-resolution spectra of Earth-like planets, including ocean worlds, snowball planets, frozen worlds, and Earth analogues, are available online, providing a tool for modellers and observers to train retrie v al algorithms, optimize search strategies, and inform models of Earth-like planets, where purple is the new green.
\end{abstract}




\section{Introduction}

With more than 5500 confirmed exoplanets and more than 30
potential Earth-like planets (exoplanets.nasa.gov, 2024 February)
our search for life in the cosmos is entering a new stage that requires including a wider diversity of life on Earth \citep{hegde2015surface, o2018vegetation, coelho2022color} and its evolution \citep{o2018vegetation, o2019expanding} to generate
our key to finding it elsewhere.

Before the emergence of chlorophyll-based oxygenic photosynthesis
\citep{lyons2014rise, hamilton2016role}, anoxygenic photosynthesis prevailed, relying on bacteriochlorophylls \citep{hohmann2011evolution}. Oxygenic photosynthesis on modern Earth is often linked to lush green landscapes. Anoxygenic phototrophs, which can inhabit both aquatic and land environments, can also colour their environments. Early Earth may have been dominated by purple-pigmented bacteria, including purple non-sulfur bacteria (PNSB) and purple sulfur bacteria (PSB) \citep{brocks2005biomarker, sanroma2013characterizing}.

PSB are phototrophic anoxygenic bacteria that prefer reduced sulfur compounds as electron donors instead of water, thus not producing oxygen as a byproduct of photosynthesis. PNSB are photoheterotrophic bacteria meaning they use their photosystems to mainly produce energy and electric potential while still requiring an external organic carbon source. PNSB are versatile organisms that can use some inorganic compounds (e.g. H$_2$ ) as electron donors, and that are less sensitive to oxygen and darkness than PSB, although more sensitive to sulfide \citep{madigan2009purple}.

\begin{figure}
    \centering
    \includegraphics[width=\columnwidth]{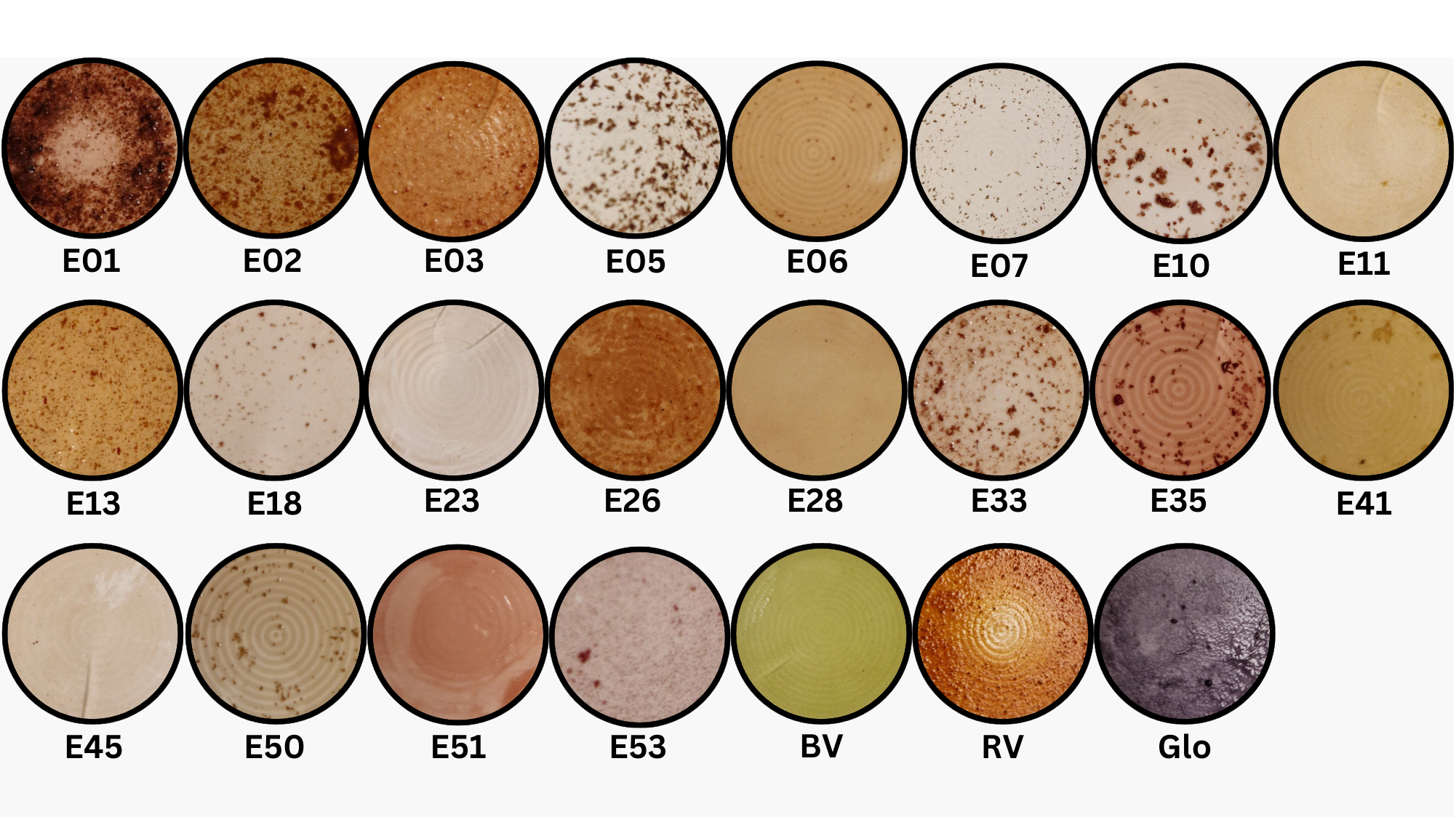}
    \vspace{-1em}
    \caption{Filters showing the measured purple non-sulfur bacteria (E01, E02, E11, E18, E23, E26, E33, E45, BV, and RV), purple sulfur bacteria (E03, E05, E06, E07, E10, E13, E28, E35, E41, E50, E51, E53), and purple cyanobacteria (Glo). For details on the biota see Table S1.}
    \label{fig:purple_non-sulfur_bacteria}
    \vspace{-1.5em}
\end{figure}

PNSB and PSB use biopigments called carotenoids as antennas to harvest light energy in the visible range \citep{magdaong2014high, ma2022aromatic}. These bacteria also use bacteriochlorophylls (BChl) \textit{a} and \textit{b}, which serve as both antenna and reaction centre pigments and can harness visible to infrared (IR) radiation, absorbing and reflecting from 750 to 1100 nm \citep{kimura2023advances}. This spectral versatility arguably provides a competitive edge, allowing organisms with BChls to thrive in diverse and often challenging ecological niches across our planet, even deep-sea hydrothermal vents \citep{beatty2005obligately}. In deep-sea conditions, they rely on IR energy to drive anoxygenic phototrophy, where radiation longer than 850 nm can serve as their standard energy source. These characteristics could allow PNSB and PSB to thrive on exoplanets orbiting M-type stars because the spectral energy distribution of M-stars overlaps with the optimal spectral range for absorption and reflectance by BChl-\textit{a} and -\textit{b} \citep[][]{wolstencroft2002photosynthesis, tinetti2006detectability, kiang2007spectral, schwieterman2018exoplanet, lehmer2021peak, duffy2023photosynthesis}{}{}.

Even Earth-analogue planets orbiting Sun-analogue stars can promote purple bacteria. \citet{sanroma2013characterizing} measured the reflectivity of a single non-sulfur bacterium, \textit{Rhodobacter sphaeroides}, and modelled the spectrum of a purple Archean Earth from it. On modern Earth, purple cyanobacteria, PNSB, and PSB, colour surface landscapes in shallow waters, coasts, and marshes \citep{guglielmi1981structure, wilbanks2014microscale}. Their colour intensifies when light becomes scarce, and pigment production increases \citep[][]{magdaong2014high}{}{}, making purple bacteria particularly intriguing candidates for exoplanetary habitability biopigment assessments. However, to date, there is no reference data base for the reflectance spectra of the diverse biopigments of purple biota.

Here we grow and measure the reflectance spectra of a wide range of PNSB, PSB, and a purple cyanobacterium (hereafter we refer to our collection as ‘purple bacteria’). We show bacteria closely related to \textit{R. sphaeroides} for reference in Figs 1 and 2 and expand on the measurement by \citet{sanroma2013characterizing} by measuring the reflectance for a wide range of purple bacteria, including \textit{Blastochloris viridis}, which broadens the range of photosynthesis to the near-IR. Until now, most models for terrestrial exoplanets that incorporate surface biota exhibit a bias towards green pigments \citep{seager2005vegetation, o2018vegetation, o2019expanding, schwieterman2018exoplanet}. This bias is primarily due to the limited reflectance spectra data for purple bacteria, which have not been extensively measured. 

In this study, we present a catalogue of reflectance spectra of biopigments for a range of purple bacteria in fresh and dry states \citep[following][]{coelho2022color}{}{}, measured with an integrating sphere to mimic exoplanet observing conditions \citep[see][]{hegde2015surface}{}{}. We used those measurements to model the spectra of a range of Earth-like planets dominated by diverse purple bacteria for Earth analogue planets, ocean worlds, snowball planets, as well as frozen worlds, with and without clouds. Our goal was to explore how their biopigments can shape the reflectance spectra of habitable worlds.

We created a data base for the reflectance spectra of biopigments of a wide range of purple bacteria from 400 to 2500 nm. This data base is available online (\url{https://doi.org/10.5281/zenodo.10697546}) as a resource for observers and modellers to inform design decisions and optimize observations. We use those measurements to simulate high-resolution reflectance spectra of Earth-like exoplanets dominated by purple bacteria providing a data set to train retrieval algorithms and inform the observing strategies for telescopes like the European Extremely Large Telescope \citep{gilmozzi2007european} and the design of upcoming telescopes like the Habitable World Observatory (HWO) \citep{vaughan2023chasing}. These data bases provide a critical resource for observers and modellers searching for life in the cosmos.

\section{Methods}
We grew a diverse set of purple bacteria in the laboratory, measured the reflectivity of their biopigments, and then used that data to simulate the reflectance spectra of Earth-like exoplanets dominated by purple biota. 

We present enrichment samples (E01 to E53) and pure cultures (RV, BV, and Glo) – please see Fig.\ref{fig:purple_non-sulfur_bacteria}. The difference between an enrichment and a pure culture is that a pure culture is a single species while an enrichment can contain many species but is typically dominated by a species of interest. Life as we know it exists in complex communities and not in pure cultures and those communities will not always be dominated by one species. Thus, our sampling relates to an exoplanet surface dominated by an ecosystem enriched in purple bacteria (a similar phenomenon to algae blooms and watermelon snow, where a single group of microbes dominates a whole landscape on modern Earth).

\subsection{Biological measurements}
All cultures and enrichments were sampled from aquatic biomes, and the specifics of their location of origin can be found in the supplementary materials. Briefly, RV and BV were obtained from S. Zinder’s laboratory at Cornell University (NY, US). PSB E53 was sampled by L.F. Coelho and S. Zinder from Beebe Lake (NY, US) and enriched using a modular medium for PSB with 3 mM of sulfide. PSB E51 and \textit{Gloeobacter violaceus} PCC 7421 were obtained from T. Hamilton’s laboratory at the University of Minnesota (MN, US). PSB E51 was enriched from a sulfidic spring in NY. PSB and PNSB enrichments (E01-E50) were courtesy of the Microbial Diversity 2023 Course (Marine Biological Laboratory, Woods Hole, MA, US). 

We grew PSB and PNSB in a modular mineral basal medium based on Pfennig’s SL9 medium \citep{tschech1984growth}. The PNSB grew under an N$_2$/CO$_2$ atmosphere with 1 g L$^{-1}$ sodium succinate as the carbon source, and the PSB grew under autotrophic conditions in a 20 per cent CO$_2$ atmosphere and 1 per cent bicarbonate. The purple cyanobacterium \textit{ G. violaceus} PCC 7421 and blue-green cyanobacterium \textit{Anabaena} sp. UTEX 2576 were grown in BG11 medium. All cultures were maintained at room temperature. Differences in the media and light requirements for each culture or enrichment can be found in the supplementary material – Table S1. Depending on the culture, the time required for growth varied from about 24 h to 1 week. A volume of 20 mL of the cultures/enrichments was transferred to 50 mL Falcon centrifuge tubes until filtration.

\subsection{Sample data collection}
We deposited cultures on to a 0.45 $\micron$ 25 mm white mixed cellulose ester filter using a 10 mL syringe until saturation was reached \citep[as previously described in][]{hegde2015surface, coelho2022color}. Subsequently, we measured hemispherical reflectance from 400--2500 nm at intervals of 1 nm, using three detectors: a 512-element silicon photodiode array (up to 1000 nm) and two indium–gallium–arsenide photodiodes (1000--1800 and 1800--2500 nm) using an ASD FieldSpec 4 Spectrometer coupled with an ASD integrating sphere \citep[as described in detail in][]{hegde2015surface, coelho2022color}. Measuring the reflectance of the biopigments using an integrating sphere mimics the observing geometry of an exoplanet \citep[as previously described in][]{hegde2015surface}.

Dry biopigments show stronger reflectance than fresh samples \citep[see also ][]{coelho2022color}{}{} and thus we measured the same sample twice: immediately after deposition on the filter (referred to as ‘fresh’) and after a week at the benchtop in the dark (referred to as ‘dry’). Filters with only culture media served as controls.

\subsection{Data generation and simulations}
We used the established Exo-Prime II model \citep[see e.g.][]{kaltenegger2020finding, madden2020surfaces} to generate high-resolution atmospheric spectra with a minimum resolution of 
$\lambda/\Delta\lambda = 100\,000$
We chose a slightly colder modern Earth for our analysis with an average surface temperature of 273 K to allow for a wide range of surface conditions, using modern Earth outgassing rates \citep[e.g.][]{rugheimer2013spectral, kaltenegger2021finding}{}{} and a 10 per cent decreased solar irradiation to account for a planet that could be covered with liquid or frozen water. This slightly colder atmosphere does not show significant changes in spectral features compared to modern Earth, except for slightly reduced water features due to the lower humidity of the atmosphere. However, these slight differences do not influence our analysis of the effect of surface biota on the spectra.

Ocean and snow surface albedo have been selected from the United States Geological Survey (USGS) Spectra Library \citep{clark2007usgs}, and the single-layer cloud albedo is based on the 20 $\micron$ MODIS model \citep{king1997cloud, AdvancesinUnderstandingCloudsfromISCCP} following previous work \citep[see e.g.][]{madden2020surfaces, pham2021color, pham2022follow}{}{}.

We resampled the high resolution of the reflectance spectra of Earth-like exoplanets and the low-resolution biota (Fig.~\ref{fig:modelled_reflectance_spectra}), cloud, ocean, and snow albedos using the Python \texttt{SPECTRES} 2.2.0 package \citep{carnall2017spectres}. We resampled the low-resolution biota, cloud, ocean, and snow albedos to R~$= 140$ (comparable to the HWO concept, Fig. S2) and R~$= 100\,000$, comparable to high-resolution instruments such as ELTs (Fig. S3). Our exoplanet models contain reflectance spectra of frozen worlds and snowball planets, Earth-analogues, and ocean worlds with a different surface coverage of purple bacteria for atmospheres with and without clouds.

\section{Results and Discussion}
\begin{figure*}
    \centering
    \vspace{-1.5em}
    \includegraphics[width=\textwidth]{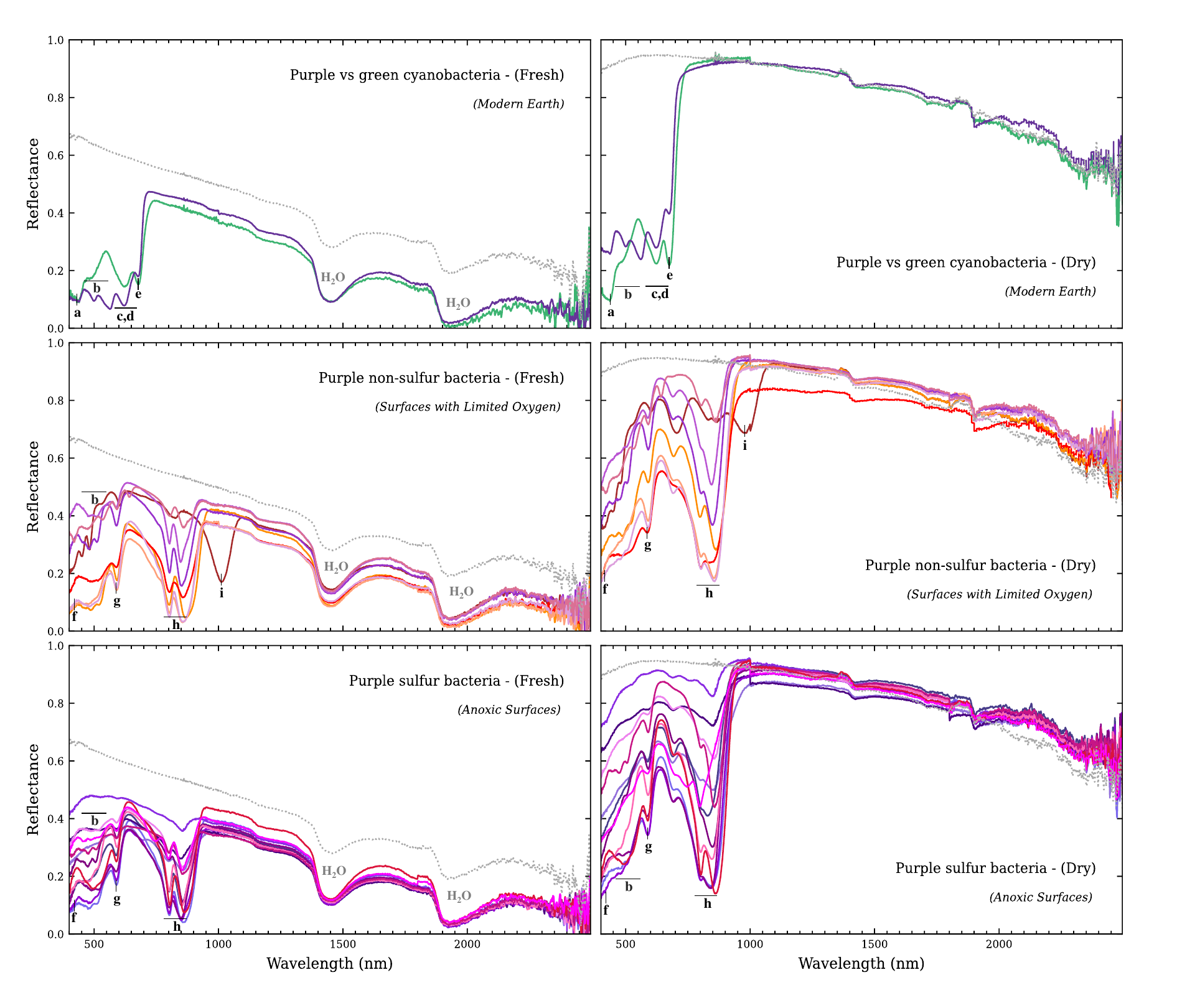}
    \vspace{-2.5em}
    \caption{Reflectance spectra of purple and blue-green cyanobacteria (top), PNSB (N~$=10$) (middle), and, PSB (N~$= 12$) (bottom) where N denotes the number of samples. Labels of main pigments identified: (a) Chl-\textit{a} Soret band, (b) Carotenoids, (c) Chl-\textit{a} Qx band, (d) Phycobilins, (e) Chl-\textit{a} Qy band, (f) BChl-\textit{a} Soret band, (g) BChl-\textit{a}, \textit{b} Qx band, (h) BChl-\textit{a} Qy band, and (i) BChl-\textit{b} Qy band. Dry samples (right) show a stronger reflectance than fresh samples (left), water absorption features at 1490 and 1900 nm are stronger in the fresh samples. The reflectance of the culture medium alone is shown as a dotted line for control.}
    \label{fig:reflectance_spectra}
\end{figure*}

\begin{figure*}
    \centering
    \includegraphics[width=\textwidth]{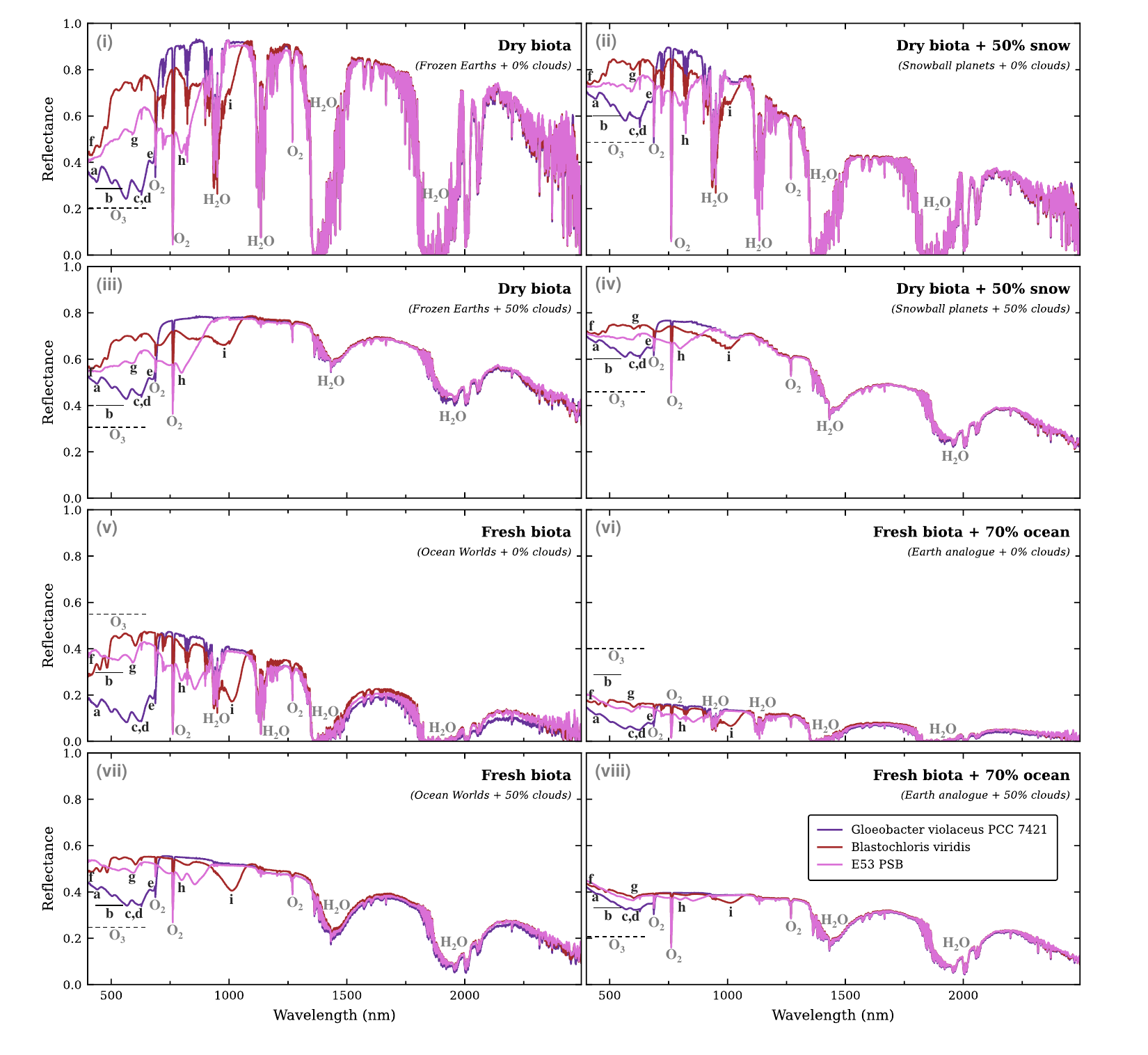}
    \vspace{-2.0em}
    \caption{Modelled reflectance spectra of Earth-like planets (top) for a frozen planet covered with 100 per cent dry biota (i) and a snowball planet covered with 50 per cent dry biota and 50 per cent snow (ii), which increases the overall reflectivity slightly. Adding a 50 per cent cloud cover (2nd row, iii and iv) reduces the overall reflectivity of these cold worlds due to the lower albedo of clouds compared to snow and the dry biota (see Fig. S1). The models of an ocean world (100 per cent fresh biota) and an Earth-analogue planet (70 per cent ocean and 30 per cent fresh biota) (3rd row, v and vi) show that the overall reflectivity of the planet decreases due to the lower reflectivity of the fresh biota as well as the low reflectivity of oceans compared to frozen worlds. Adding clouds to the planet (bottom, vii and viii) increases the overall reflectivity of the models because clouds reflect more light than oceans or fresh biota. Labels: (a) Chl-\textit{a} Soret band, (b) Carotenoids, (c) Chl-\textit{a} Qx band, (d) Phycobilins, (e) Chl-\textit{a} Qy band, (f) BChl-\textit{a} Soret band, (g) BChl-\textit{a}, \textit{b} Qx band, (h) BChl-\textit{a} Qy band, and (i) BChl-\textit{b} Qy band (Table S2). Atmospheric absorption features: O$_3$ (450 to 740 nm), O$_2$ (690, 760, and 1260 nm), and H$_2$O (950, 1150, 1490, and 1900 nm).}
    \vspace{0.5em}
    \label{fig:modelled_reflectance_spectra}
\end{figure*}

To expand our baseline for finding life in the cosmos, we have measured the reflectance of purple bacteria that thrive across a range of anoxic and oxic environments. We present the samples in Fig.~\ref{fig:purple_non-sulfur_bacteria}, the reflectance of the biopigments in Fig.~\ref{fig:reflectance_spectra}, and the model reflectance spectra of Earth-like exoplanets dominated by a range of such purple biota in Fig.~\ref{fig:modelled_reflectance_spectra}.

In Fig.~\ref{fig:purple_non-sulfur_bacteria}, we show 23 purple bacteria samples, originating from oxic to anoxic conditions. Despite being generally referred to as ‘purple’, these organisms can exhibit a wide range of colours as shown
here, including yellow, orange, brown, or even red due to the presence of different carotenoids. Fig.~\ref{fig:reflectance_spectra} shows the distinct reflectivity of the 23 purple bacteria measured in both fresh and dry conditions. The main features can be identified in the spectra range between 400--1100 nm. Our spectra show a minimum resolution of 125 comparable to the intended resolution of the Habitable World Observatory  (R~$= 140$). 
We resampled the data to R~$= 140$ (Fig. S2) showing no significant changes in the biota signatures. Fig.~\ref{fig:reflectance_spectra} (top) shows the reflectance spectra of the purple cyanobacterium \textit{G. violaceus} compared to that of \textit{Anabaena} sp., a typical blue-green cyanobacterium, both of which thrive in oxygen-rich environments. Despite inhabiting similar biomes, the purple-coloured cyanobacterium \textit{G. violaceus} exhibits a distinct reflectance spectrum. The Anabaena sp. reflectance spectrum is characterized by a prominent Chlorophyll \textit{a} (Chl-\textit{a}) peak at 548 nm (see Table S2), a typical reflectance signature of organisms with a high Chl-\textit{a} content \citep[see][Fig.~\ref{fig:reflectance_spectra}]{coelho2022color}{}{}. Both cyanobacteria have Chl-\textit{a} absorption features at 437--439 and 675--677 nm and a phycobilin absorption feature at 620--624 nm, typical of cyanobacteria. Purple cyanobacteria produce distinct carotenoids (440--550 nm), \textit{G. violaceus} spectrum (Fig.~\ref{fig:reflectance_spectra} top) displays distinct carotenoid absorption features at 500 and 560 nm (visible in dry samples) in the spectrum. Fig.~\ref{fig:reflectance_spectra} (top) also shows some overlaps of biopigment features between the biota: echinenone, a carotenoid usually present in \textit{Gloeobacter} absorbs at 440 nm and is similar to the Soret band of Chl-\textit{a}. Similarly, \textit{Anabaena} sp. presents an absorption feature at 488 nm that agrees with the reference peak of canthaxanthin, another carotenoid \citep{takaichi2005myxol}.

In addition to oxic environments (Fig.~\ref{fig:reflectance_spectra} top), we show the reflectivity of purple bacteria that grow under different conditions: PNSB adapted to environments with limited oxygen (Fig.~\ref{fig:reflectance_spectra} middle) and PSB in anoxic conditions (Fig.~\ref{fig:reflectance_spectra} bottom). In both panels part of the BChl Soret bands can be seen at 400–420 nm. In PSB and PNSB, BChl-\textit{a} and -\textit{b} Qy bands are present in the near-IR (800--1100 nm). The BChl-\textit{a} spectral Q-band features exhibit minimal variability among organisms, with absorption features at 800--805 nm and 850–855 nm typical of PNSB and PSB \citep{kimura2023advances}. BChl \textit{b} displays a considerably different Q band, with the position of the peaks ranging from 820 to 1100 nm. Note that Fig.~\ref{fig:reflectance_spectra} (middle) displays a PNSB (\textit{B. viridis}) that absorbs at 1010 nm and can use this energy for phototrophy, suggesting that ecosystems dominated by purple biota may be more likely to exist on planets orbiting cooler M-stars that provide stellar energy distribution peaking at longer, redder, wavelengths \citep[as suggested previously by][]{wolstencroft2002photosynthesis, tinetti2006detectability, kiang2007spectral, lehmer2021peak}. See Table S2 for detailed listing of the biopigments spectral signatures in purple bacteria.

We then simulated the reflectance spectra of Earth-like planets shown in Fig.~\ref{fig:modelled_reflectance_spectra} for Earth-analogue planets (70 per cent ocean and 30 per cent land covered with fresh biota), ocean worlds (covered with 100 per cent fresh biota), frozen worlds (covered with 100 per cent dry biota) and snowball worlds(covered with 50 per cent dry biota and 50 per cent snow). We then added 50 per cent cloud cover to show the effect of clouds on the reflectance spectra (see Fig. S1 for the albedos of snow, oceans, clouds, and the selected fresh and dry biota). Models of the reflectance spectra of Earth-like exoplanets show a strong reflectivity for planetary models covered in dry biota and snow, but lower reflectivity for planets with open oceans and fresh biota (see Fig.~\ref{fig:modelled_reflectance_spectra}). A snowball planet with 50 per cent dry biota and 50 per cent snow (Fig.~\ref{fig:modelled_reflectance_spectra}–ii) shows the highest overall reflectivity due to the high reflectivity of snow and dry biota (see Fig. S1).

The frozen planet models, covered with 100 per cent dry biota, show the second highest reflectivity, only slightly lower than the snowball planet. Adding a 50 per cent cloud cover (Fig.~\ref{fig:modelled_reflectance_spectra}–iii and –iv) reduces the overall reflectivity of both models slightly due to the lower albedo of clouds compared to snow and the dry biota. For Earth-analog models of a planet (70 per cent ocean and 30 per cent land covered by fresh biota) and ocean worlds (100 per cent fresh biota) (Fig.~\ref{fig:modelled_reflectance_spectra}–v and –vi) the overall reflectivity is lower than for the previous cases due to the lower reflectivity of the fresh biota compared to the dry biota and the low reflectivity of oceans compared to snow. Adding clouds to the models (bottom) increases the overall reflectivity slightly (Fig.~\ref{fig:modelled_reflectance_spectra}–vii and –viii).

There is an overlap between biopigments and atmospheric absorption features. B(Chl)-a Qy absorption features (680 nm for Chl-\textit{a} and 800--850 nm for BChl-\textit{a}) are too close, in the lower and upper limit of O$_2$ (760 nm). Note that BChl-\textit{a} Qy of PSB (anoxic surfaces) has been reported to also absorb at 900--963 nm \citep{kimura2023advances}, which does not overlap with O$_2$, but may overlap with an intense H$_2$O feature at 950 nm. Similarly, BChl-\textit{b} Qy absorption features far in the 1000 nm may overlap with the same H2O feature. Carotenoids overlap with O$_3$ (450--740 nm) although in Fig.~\ref{fig:modelled_reflectance_spectra} that does not significantly mask the biopigments features.

\citet{barrientos2023search} showed changes in surface albedos provide vital and usable information for observations, even with a low signal-to-noise ratio. Their quantitative work was applied to a general ‘edge feature’ in the spectrum, based on the specific biological surface albedo caused by green photosynthetic organisms – the ‘red edge’ \citep{barrientos2023search}.

Traditionally, the ‘red edge’ \citep[e.g.][]{sagan1993search, seager2005vegetation, arnold2008earthshine, o2018vegetation, o2019expanding}{}{} has been used as a surface biosignature for modern Earth (see Fig.~\ref{fig:reflectance_spectra}, top panel). However, a photosynthesis ‘red edge’ is an advanced surface biosignature for life on Earth that may only have become detectable between 0.75 and 1.2 billion years ago \citep{o2018vegetation}. Sharp changes in the surface albedo in the visible range could have a similar observation promise \citep{barrientos2023search}. These results stress the importance of applying wavelength-dependent surface features to retrieval models to estimate atmospheric features such as H$_2$O, O$_2$, and O$_3$ correctly. Future studies applying a similar quantitative method will determine the detectability of purple bacteria features.

Here we extend our search for surface biosignatures to a much wider range of terrestrial exoplanet environments using biopigments of purple bacteria. Further, we expand the type of habitats such biota could dominate including exoplanets with anoxic or microaerobic environments. Some spectral features of (B)Chl and carotenoids lie close to atmospheric features like H$_2$O and O$_3$; thus, the community needs to consider a diverse set of surface biopigments in the retrieval models \citep{barrientos2023search}, not only to detect signs of life but also to retrieved atmospheric features co-existing in the same wavelengths as surface features.

In this study, we focused on terrestrial Earth-like planets. Other types of proposed exoplanet environments, like Hycean planets, covered by oceans under substantial H$_2$ atmospheres \citep{madhusudhan2021habitability} could also be promising candidates for purple bacterial ecosystems. Some purple bacteria organisms can use H$_2$ as a primary electron donor. Thus, the abundance of H$_2$ on these proposed planets may facilitate biomass production.

The distinct reflectivity of the biopigments of purple bacteria offers surface biosignatures for a wide range of conditions -- from oxic to anoxic conditions -- to the point of detection with upcoming telescopes like ELTs or the Habitable World Telescope.

\section{Conclusions}
Using life on Earth as our guide we identified purple bacteria as biota that could dominate a wide range of environments on exoplanets. Especially on exoplanets orbiting M-stars, purple bacteria could arguably be dominant because they can use red to infrared light to fuel photosynthesis. Where chlorophyll-based photosynthesis thrives under the Sun’s spectral energy distribution, a range of purple bacteria could thrive under a red star’s light.

We provide a reflectance spectra data base of the biopigments of purple bacteria as a resource for modelers and observers to optimize search strategies for life with upcoming telescopes. We expand the range of surface biosignatures on potentially habitable exoplanets beyond the ‘red edge’ traditionally used to search for vegetation. Despite being generally referred to as purple bacteria, these organisms can exhibit a wide range of colours, including yellow, orange, brown, or red due to the presence of different carotenoids, and could colour the surface of exoplanets in many shades.

Based on these reflectance measurements we created an exoplanet data base of high-resolution reflectance spectra for Earth-like exoplanets including ocean worlds, Earth-analogues, frozen worlds, and snowball planets to explore the effects of purple bacteria on their overall reflectance spectra.

Our models show that depending on the surface coverage of the biota and the cloud coverage, a wide variety of terrestrial planets could show signs of purple bacteria surface biopigments. While it is unknown whether life -- or purple bacteria -- can evolve on other worlds, purple might just be the new green in the search for surface life.

\section*{Acknowledgements}

LFC acknowledges the funding from Fulbright Schuman. LFC and LK acknowledge the funding from the Binson Foundation. TLH acknowledges funding from NSF award 1939303. The authors acknowledge the 2023 Microbial Diversity course from the Marine Biological Laboratory with funding from the Moore Foundation, Simons Foundation, US Department of Energy, and NSF award 22055876 (PIs: RachelWitaker, Scott Dawson, and L. Hyman) for the samples of PNSB and PSB (E01 to E50). The authors expressly acknowledge the Microbial Diversity class of 2023, including Isabelle (Izzy) Lakis, Priyanka Chatterjee, Amaranta Khan, Etan Dieppa, Nicole Dames, Maria Jose Rodriguez, (Mache), Lucy Androsuik, McKenzie Powers, Becky Parales, Huda Usman, Marwa Baloza, and Allison Higgins for the enrichments E02-E50, respectively. The authors acknowledge Jeff Havig for his assistance in collecting sample E51. The authors acknowledge Adam Langeveld and Rebecca Payne for the very useful modelling insights.

\section*{Data Availability}
These data bases are available online (\url{https://doi.org/10.5281/zenodo.10697546}).



\bibliographystyle{mnras}
\bibliography{references} 

\begin{thebibliography}{}
\makeatletter
\relax
\def\mn@urlcharsother{\let\do\@makeother \do\$\do\&\do\#\do\^\do\_\do\%\do\~}
\def\mn@doi{\begingroup\mn@urlcharsother \@ifnextchar [ {\mn@doi@} {\mn@doi@[]}}
\def\mn@doi@[#1]#2{\def\@tempa{#1}\ifx\@tempa\@empty \href {http://dx.doi.org/#2} {doi:#2}\else \href {http://dx.doi.org/#2} {#1}\fi \endgroup}
\def\mn@eprint#1#2{\mn@eprint@#1:#2::\@nil}
\def\mn@eprint@arXiv#1{\href {http://arxiv.org/abs/#1} {{\tt arXiv:#1}}}
\def\mn@eprint@dblp#1{\href {http://dblp.uni-trier.de/rec/bibtex/#1.xml} {dblp:#1}}
\def\mn@eprint@#1:#2:#3:#4\@nil{\def\@tempa {#1}\def\@tempb {#2}\def\@tempc {#3}\ifx \@tempc \@empty \let \@tempc \@tempb \let \@tempb \@tempa \fi \ifx \@tempb \@empty \def\@tempb {arXiv}\fi \@ifundefined {mn@eprint@\@tempb}{\@tempb:\@tempc}{\expandafter \expandafter \csname mn@eprint@\@tempb\endcsname \expandafter{\@tempc}}}

\bibitem[\protect\citeauthoryear{Arnold}{Arnold}{2008}]{arnold2008earthshine}
Arnold L.,  2008, Strategies of Life Detection, pp 323--333

\bibitem[\protect\citeauthoryear{Barrientos, MacDonald, Lewis  \& Kaltenegger}{Barrientos et~al.}{2023}]{barrientos2023search}
Barrientos J.~G.,  MacDonald R.~J.,  Lewis N.~K.,   Kaltenegger L.,  2023, The Astrophysical Journal, 946, 96

\bibitem[\protect\citeauthoryear{Beatty et~al.,}{Beatty et~al.}{2005}]{beatty2005obligately}
Beatty J.~T.,  et~al., 2005, Proceedings of the National Academy of Sciences, 102, 9306

\bibitem[\protect\citeauthoryear{Brocks, Love, Summons, Knoll, Logan  \& Bowden}{Brocks et~al.}{2005}]{brocks2005biomarker}
Brocks J.~J.,  Love G.~D.,  Summons R.~E.,  Knoll A.~H.,  Logan G.~A.,   Bowden S.~A.,  2005, Nature, 437, 866

\bibitem[\protect\citeauthoryear{Carnall}{Carnall}{2017}]{carnall2017spectres}
Carnall A.,  2017, arXiv preprint arXiv:1705.05165

\bibitem[\protect\citeauthoryear{Clark, Swayze, Wise, Livo, Hoefen, Kokaly  \& Sutley}{Clark et~al.}{2007}]{clark2007usgs}
Clark R.~N.,  Swayze G.~A.,  Wise R.~A.,  Livo K.~E.,  Hoefen T.~M.,  Kokaly R.~F.,   Sutley S.~J.,  2007, Technical report, USGS digital spectral library splib06a.
US Geological Survey

\bibitem[\protect\citeauthoryear{Coelho et~al.,}{Coelho et~al.}{2022}]{coelho2022color}
Coelho L.~F.,  et~al., 2022, Astrobiology, 22, 313

\bibitem[\protect\citeauthoryear{Duffy, Canchon, Haworth, Gillen, Chitnavis  \& Mullineaux}{Duffy et~al.}{2023}]{duffy2023photosynthesis}
Duffy C.~D.,  Canchon G.,  Haworth T.~J.,  Gillen E.,  Chitnavis S.,   Mullineaux C.~W.,  2023, Monthly Notices of the Royal Astronomical Society, 526, 2265

\bibitem[\protect\citeauthoryear{Gilmozzi \& Spyromilio}{Gilmozzi \& Spyromilio}{2007}]{gilmozzi2007european}
Gilmozzi R.,  Spyromilio J.,  2007, The Messenger, 127, 3

\bibitem[\protect\citeauthoryear{Guglielmi, Cohen-Bazire  \& Bryant}{Guglielmi et~al.}{1981}]{guglielmi1981structure}
Guglielmi G.,  Cohen-Bazire G.,   Bryant D.~A.,  1981, Archives of microbiology, 129, 181

\bibitem[\protect\citeauthoryear{Hamilton, Bryant  \& Macalady}{Hamilton et~al.}{2016}]{hamilton2016role}
Hamilton T.~L.,  Bryant D.~A.,   Macalady J.~L.,  2016, Environmental microbiology, 18, 325

\bibitem[\protect\citeauthoryear{Hegde, Paulino-Lima, Kent, Kaltenegger  \& Rothschild}{Hegde et~al.}{2015}]{hegde2015surface}
Hegde S.,  Paulino-Lima I.~G.,  Kent R.,  Kaltenegger L.,   Rothschild L.,  2015, Proceedings of the National Academy of Sciences, 112, 3886

\bibitem[\protect\citeauthoryear{Hohmann-Marriott \& Blankenship}{Hohmann-Marriott \& Blankenship}{2011}]{hohmann2011evolution}
Hohmann-Marriott M.~F.,  Blankenship R.~E.,  2011, Annual review of plant biology, 62, 515

\bibitem[\protect\citeauthoryear{Kaltenegger \& Lin}{Kaltenegger \& Lin}{2021}]{kaltenegger2021finding}
Kaltenegger L.,  Lin Z.,  2021, The Astrophysical Journal Letters, 909, L2

\bibitem[\protect\citeauthoryear{Kaltenegger, Lin  \& Rugheimer}{Kaltenegger et~al.}{2020}]{kaltenegger2020finding}
Kaltenegger L.,  Lin Z.,   Rugheimer S.,  2020, The Astrophysical Journal, 904, 10

\bibitem[\protect\citeauthoryear{Kiang, Siefert, Govindjee  \& Blankenship}{Kiang et~al.}{2007}]{kiang2007spectral}
Kiang N.~Y.,  Siefert J.,  Govindjee  Blankenship R.~E.,  2007, Astrobiology, 7, 222

\bibitem[\protect\citeauthoryear{Kimura, Tani, Madigan  \& Wang-Otomo}{Kimura et~al.}{2023}]{kimura2023advances}
Kimura Y.,  Tani K.,  Madigan M.~T.,   Wang-Otomo Z.-Y.,  2023, The Journal of Physical Chemistry B, 127, 6

\bibitem[\protect\citeauthoryear{King, Tsay, Platnick, Wang  \& Liou}{King et~al.}{1997}]{king1997cloud}
King M.~D.,  Tsay S.-C.,  Platnick S.~E.,  Wang M.,   Liou K.-N.,  1997, MODIS Algorithm Theoretical Basis Document, 1997, 440

\bibitem[\protect\citeauthoryear{Lehmer, Catling, Parenteau, Kiang  \& Hoehler}{Lehmer et~al.}{2021}]{lehmer2021peak}
Lehmer O.~R.,  Catling D.~C.,  Parenteau M.~N.,  Kiang N.~Y.,   Hoehler T.~M.,  2021, Frontiers in Astronomy and Space Sciences, 8, 689441

\bibitem[\protect\citeauthoryear{Lyons, Reinhard  \& Planavsky}{Lyons et~al.}{2014}]{lyons2014rise}
Lyons T.~W.,  Reinhard C.~T.,   Planavsky N.~J.,  2014, Nature, 506, 307

\bibitem[\protect\citeauthoryear{Ma \& Cui}{Ma \& Cui}{2022}]{ma2022aromatic}
Ma J.,  Cui X.,  2022, Geosystems and Geoenvironment, 1, 100045

\bibitem[\protect\citeauthoryear{Madden \& Kaltenegger}{Madden \& Kaltenegger}{2020}]{madden2020surfaces}
Madden J.,  Kaltenegger L.,  2020, Monthly Notices of the Royal Astronomical Society, 495, 1

\bibitem[\protect\citeauthoryear{Madhusudhan, Piette  \& Constantinou}{Madhusudhan et~al.}{2021}]{madhusudhan2021habitability}
Madhusudhan N.,  Piette A.~A.,   Constantinou S.,  2021, The Astrophysical Journal, 918, 1

\bibitem[\protect\citeauthoryear{Madigan, Jung, Hunter, Daldal, Thurnauer  \& Beatty}{Madigan et~al.}{2009}]{madigan2009purple}
Madigan M.,  Jung D.,  Hunter C.,  Daldal F.,  Thurnauer M.~C.,   Beatty J.,  2009, Advances in photosynthesis and respiration. Dordrecht: Springer

\bibitem[\protect\citeauthoryear{Magdaong et~al.,}{Magdaong et~al.}{2014}]{magdaong2014high}
Magdaong N.~M.,  et~al., 2014, The Journal of Physical Chemistry B, 118, 11172

\bibitem[\protect\citeauthoryear{O'Malley-James \& Kaltenegger}{O'Malley-James \& Kaltenegger}{2018}]{o2018vegetation}
O'Malley-James J.~T.,  Kaltenegger L.,  2018, Astrobiology, 18, 1123

\bibitem[\protect\citeauthoryear{O’Malley-James \& Kaltenegger}{O’Malley-James \& Kaltenegger}{2019}]{o2019expanding}
O’Malley-James J.~T.,  Kaltenegger L.,  2019, The Astrophysical Journal Letters, 879, L20

\bibitem[\protect\citeauthoryear{Pham \& Kaltenegger}{Pham \& Kaltenegger}{2021}]{pham2021color}
Pham D.,  Kaltenegger L.,  2021, Monthly Notices of the Royal Astronomical Society, 504, 6106

\bibitem[\protect\citeauthoryear{Pham \& Kaltenegger}{Pham \& Kaltenegger}{2022}]{pham2022follow}
Pham D.,  Kaltenegger L.,  2022, Monthly Notices of the Royal Astronomical Society: Letters, 513, L72

\bibitem[\protect\citeauthoryear{Rossow \& Schiffer}{Rossow \& Schiffer}{1999}]{AdvancesinUnderstandingCloudsfromISCCP}
Rossow W.~B.,  Schiffer R.~A.,  1999, \mn@doi [Bulletin of the American Meteorological Society] {10.1175/1520-0477(1999)080<2261:AIUCFI>2.0.CO;2}, 80, 2261

\bibitem[\protect\citeauthoryear{Rugheimer, Kaltenegger, Zsom, Segura  \& Sasselov}{Rugheimer et~al.}{2013}]{rugheimer2013spectral}
Rugheimer S.,  Kaltenegger L.,  Zsom A.,  Segura A.,   Sasselov D.,  2013, Astrobiology, 13, 251

\bibitem[\protect\citeauthoryear{Sagan, Thompson, Carlson, Gurnett  \& Hord}{Sagan et~al.}{1993}]{sagan1993search}
Sagan C.,  Thompson W.~R.,  Carlson R.,  Gurnett D.,   Hord C.,  1993, Nature, 365, 715

\bibitem[\protect\citeauthoryear{Sanrom{\'a}, Pall{\'e}, Parenteau, Kiang, Guti{\'e}rrez-Navarro, L{\'o}pez  \& Monta{\~n}{\'e}s-Rodr{\'\i}guez}{Sanrom{\'a} et~al.}{2013}]{sanroma2013characterizing}
Sanrom{\'a} E.,  Pall{\'e} E.,  Parenteau M.,  Kiang N.,  Guti{\'e}rrez-Navarro A.,  L{\'o}pez R.,   Monta{\~n}{\'e}s-Rodr{\'\i}guez P.,  2013, The Astrophysical Journal, 780, 52

\bibitem[\protect\citeauthoryear{Schwieterman et~al.,}{Schwieterman et~al.}{2018}]{schwieterman2018exoplanet}
Schwieterman E.~W.,  et~al., 2018, Astrobiology, 18, 663

\bibitem[\protect\citeauthoryear{{Seager}, {Turner}, {Schafer}  \& {Ford}}{{Seager} et~al.}{2005}]{seager2005vegetation}
{Seager} S.,  {Turner} E.~L.,  {Schafer} J.,   {Ford} E.~B.,  2005, \mn@doi [Astrobiology] {10.1089/ast.2005.5.372}, \href {https://ui.adsabs.harvard.edu/abs/2005AsBio...5..372S} {5, 372}

\bibitem[\protect\citeauthoryear{Takaichi, Mochimaru, Maoka  \& Katoh}{Takaichi et~al.}{2005}]{takaichi2005myxol}
Takaichi S.,  Mochimaru M.,  Maoka T.,   Katoh H.,  2005, Plant and cell physiology, 46, 497

\bibitem[\protect\citeauthoryear{Tinetti, Rashby  \& Yung}{Tinetti et~al.}{2006}]{tinetti2006detectability}
Tinetti G.,  Rashby S.,   Yung Y.~L.,  2006, The Astrophysical Journal, 644, L129

\bibitem[\protect\citeauthoryear{Tschech \& Pfennig}{Tschech \& Pfennig}{1984}]{tschech1984growth}
Tschech A.,  Pfennig N.,  1984, Archives of Microbiology, 137, 163

\bibitem[\protect\citeauthoryear{Vaughan et~al.,}{Vaughan et~al.}{2023}]{vaughan2023chasing}
Vaughan S.~R.,  et~al., 2023, Monthly Notices of the Royal Astronomical Society, 524, 5477

\bibitem[\protect\citeauthoryear{Wilbanks et~al.,}{Wilbanks et~al.}{2014}]{wilbanks2014microscale}
Wilbanks E.~G.,  et~al., 2014, Environmental microbiology, 16, 3398

\bibitem[\protect\citeauthoryear{Wolstencroft \& Raven}{Wolstencroft \& Raven}{2002}]{wolstencroft2002photosynthesis}
Wolstencroft R.,  Raven J.~A.,  2002, Icarus, 157, 535

\makeatother
\end{thebibliography}



\section*{Supporting Information}
Supplementary data are available at MNRAS online. Please note: Oxford University Press is not responsible for the content or functionality of any supporting materials supplied by the authors. Any queries (other than missing material) should be directed to the corresponding author for the article.






\bsp	
\label{lastpage}
\end{document}